\documentclass[fleqn,usenatbib]{mnras}

\usepackage{newtxtext,newtxmath}

\usepackage[T1]{fontenc}

\DeclareRobustCommand{\VAN}[3]{#2}
\let\VANthebibliography\thebibliography
\def\thebibliography{\DeclareRobustCommand{\VAN}[3]{##3}\VANthebibliography}

\newcommand{\se}{{\sc SExtractor\ }}

\newcommand{\dos}{{\sc SExtractor+PSFEx\ }}

\usepackage{graphicx}	
\usepackage{amsmath}	
\usepackage{hyperref}
\usepackage{xcolor}

\usepackage{tablefootnote}
\usepackage{tikz,xcolor,hyperref}

\definecolor{lime}{HTML}{A6CE39}
\DeclareRobustCommand{\orcidicon}{
	\begin{tikzpicture}
	\draw[lime, fill=lime] (0,0) 
	circle [radius=0.16] 
	node[white] {{\fontfamily{qag}\selectfont \tiny ID}};
	\draw[white, fill=white] (-0.0625,0.095) 
	circle [radius=0.007];
	\end{tikzpicture}
	\hspace{-2mm}
}

\foreach \x in {A, ..., Z}{\expandafter\xdef\csname orcid\x\endcsname{\noexpand\href{https://orcid.org/\csname orcidauthor\x\endcsname}
			{\noexpand\orcidicon}}
}

\title[AGN candidates in the VVV NIRGC]{AGN candidates in the VVV near-IR galaxy catalogue}

\author[Baravalle et al.]
{
\parbox[t]{\textwidth}{Laura D. Baravalle\orcidA$^{1,2}$, Eduardo O. Schmidt\orcidB$^{1,2}$, M. Victoria Alonso\orcidD$^{1,2}$, Ana Pichel\orcidE$^{4}$, Dante Minniti\orcidF$^{5,6}$,  Adriana R. Rodr\'iguez-Kamenetzky\orcidC$^{1}$, Nicola Masetti\orcidG$^{7,5}$,
Carolina Villalon\orcidH$^{1}$, Leigh C. Smith\orcidI$^{8}$ \&  Philip W. Lucas\orcidJ$^{9}$}
\vspace*{6pt} \
\\
$^{1}$ Instituto de Astronom\'ia Te\'orica y Experimental (IATE, CONICET-UNC). C\'ordoba, Argentina.\\
$^{2}$ Observatorio Astron\'omico de C\'ordoba, Universidad Nacional de C\'ordoba, Laprida 854, X5000BGR, C\'ordoba, Argentina.\\
$^{4}$ Instituto de Astronom\'ia y F\'isica del Espacio, CONICET–UBA, Argentina.\\
$^{5}$ Instituto de Astrof\'isica, Facultad de Ciencias Exactas, Universidad Andr\'es Bello, Av. Fernandez Concha 700, Las Condes, Santiago, Chile.\\
$^{6}$ Vatican Observatory, V00120 Vatican City State, Italy\\
$^{7}$ INAF - Osservatorio di Astrofisica e Scienza dello Spazio, via Piero Gobetti 101, I-40129 Bologna, Italy.\\
$^{8}$ Institute of Astronomy, University of Cambridge, Madingley Road, Cambridge CB3 0HA, UK.\\
$^{9}$ Centre for Astrophysics, University of Hertfordshire, College Lane, Hatfield AL10 9AB, UK.\\
}

\date{Accepted XXX. Received YYY; in original form ZZZ}

\pubyear{2015}

\begin{document}
\label{firstpage}
\pagerange{\pageref{firstpage}--\pageref{lastpage}}
\maketitle

\begin{abstract}

The goal of this work is to search for Active Galactic Nuclei (AGN) in the Galactic disc at very low latitudes with |b| $<$ 2$^\circ$. For this, we studied the five sources from the VVV near-infrared galaxy catalogue that have also WISE counterparts and present variability in the VIrac VAriable Classification Ensemble (VIVACE) catalogue. 
In the near-infrared colour-colour diagrams, these objects have in general redder colours compared to the rest of the sources in the field. 
In the mid-infrared ones, they are located in the AGN region, however there is a source that presents the highest interstellar extinction and different mid-IR colours to be a young stellar object (YSO). 
We also studied the source variability using two different statistical methods. The fractional variability amplitude $\sigma_{rms}$ ranges from 12.6 to 33.8, being in concordance with previous results found for type-1 AGNs. 
The slopes of the light curves are in the range (2.6$-$4.7) $\times 10^{-4}$ mag/day, also in agreement with results reported on quasars variability. 
The combination of all these results suggest that four galaxies are type-1 AGN candidates
whereas the fifth source likely a YSO candidate. 
\end{abstract}




\section{Introduction}

Galaxies beyond the Milky Way are difficult to detect and identify due to the obscuring effects of dust and high stellar densities at low Galactic latitudes (e.g., \citealt{2000A&ARv..10..211K, Baravalle2021}). 
However, the study of hidden galaxies is of great importance due to they might help to identify large-scale structures beyond our Galaxy (e.g.,\citealt{Huchra2012, Macri2019}), and obtain a more complete luminosity and density distributions  (e.g., \citealt{Peebles1980, Klypin1993}). 
In this context, the near-infrared (NIR) observations are most suitable to probe these highly obscured objects. While foreground source density is higher in the NIR, reduced foreground extinction reveals galaxies that are invisible at optical wavelengths.

In this regard, the VISTA Variables in the Vía Láctea Survey (VVV, \citealt{Minniti2010}), offers an unprecedented opportunity to discover and study new different kind of objects that have remained hidden behind the Galactic plane. 
In the last few years considerable effort has been made in the detection of extragalactic sources using this survey. 
The first work on searching for extragalactic sources was made by \cite{Amores2012} from visual inspection on specific regions. Then, \cite{Baravalle2018} proposed a methodology that used a combination of morphological and photometric parameters to detect galaxies. Using this procedure, \cite{Baravalle2021} generated the VVV near-IR galaxy catalogue (VVV NIRGC) that covers the regions of the Sourthern Galactic disc.  Also, \cite{Coldwell2014,Baravalle2019} and \cite{Galdeano2021,Galdeano2022,Galdeno2023} used similar procedure to study the near-infrared properties of galaxies in galaxy clusters.
These works were particularly focused on the galaxy detections. 
\cite{Pichel2020} made the first attempt to classify four galaxies obscured by the Milky Way as blazar candidates combining near and mid-infrared data analysis. 

Active Galactic Nuclei (AGN) are among the most energetic phenomena known in the Universe. Since their discovery several decades ago, studies at different frequencies have revealed the richness of AGN phenomenology observed from radio to $\gamma$-rays, giving rise to a vast and enthralling zoo of classifications. A full overview of AGN multi-wavelength properties can be found in \cite{Padovani2017}, who also discussed the current understanding of the unification model and highlight some open questions. They emphasise the importance of the increasingly larger and deeper sky surveys at all wavelengths for major breakthroughs in the understanding of this AGN phenomenon.

There are different classes of AGNs such as type-1 and type-2 AGNs, blazars (BL Lac, \citealt{Falomo2014}) and flat spectrum radio quasars (FSRQ), among many others. 
According to the unified model \citep[][]{Antonucci1993}, different classes of AGNs would be the same type of object, mainly depending on the orientation with respect to the observer. The central engine would in all cases be a black hole accreting matter, and particularly type-1 AGNs, i.e. quasars (QSOs) and Seyfert 1 galaxies, usually  present black holes with masses $\gtrsim$10$^6$M$_{\sun}$ \citep[e.g.,][]{Ferrarese2000,Bentz2015, Schmidt2021, Prince2022}.
In addition, AGNs usually exhibit variability in the emission \citep[e.g.,][]{Nandra1997, Edelson2002, Sandrinelli2014, Pichel2020, Husemann2022}. This variability is different according to the type of AGNs and in general it is more pronounced and with higher amplitudes in blazars than in type-1 AGNs \citep[e.g.,][]{Mao2021}.

Over the past years, the population of known AGNs has grown significantly with new surveys and catalogues \citep[e.g.,][]{Veron2010, Rembold2017, Koss2017, Donacimento2019, Husemann2022}; however, the number of AGNs at lower Galactic latitudes behind the Galactic disc remains limited due to the obscuring effects typical of these dense regions \citep[][]{Edelson2012, Pichel2020}. Recently, \cite{Fu2021, Fu2022} studied the regions of the Galactic bulge searching for QSOs at lower latitudes. They used machine learning techniques to find candidates and confirmed 204 QSOs at |b| $<$ 20$^\circ$ based on spectroscopic measurements.

In this work, we present a study of AGN candidates using near- and mid-IR data in the crowded regions of the southern Galactic disc at very low Galactic latitudes (|b| $<$ 2$^\circ$).
We analyse their colour-colour diagrams and variability; and apply different classification criteria in order to understand for the first time the nature of these AGN candidates. 
The paper is organised as follows.
Section~\ref{sec:Data} presents the observational data used in this work and 
Section~\ref{sec:variability}, the variability analysis.  
Section~\ref{sec:final} presents the discussion and final remarks.

\section{Observational Data}\label{sec:Data}

\subsection{Near- and mid-IR data}

In order to carry out this work, we have used the VVV near-IR galaxy catalogue (VVV NIRGC; \citealt{Baravalle2021}), 
the VIrac VAriable Classification Ensemble (VIVACE; \citealt{Molnar2021}) and the Wide-field Infrared Survey Explorer mission (WISE; \citealt{Wright2010}).

The VVV NIRGC is the catalogue of 5563 galaxies behind the Southern Galactic disc. 
These galaxies were detected using \dos \citep{Bertin2011} on the $J$, $H$ and $K_{s}$ images from the VVV survey. They applied different morphological constraints: \textit{CLASS\_STAR}~$< 0.3$; the SPREAD\_MODEL ($\Phi$) parameter, $\Phi> 0.002$; the radius that contains 50\% of the total flux of an object ($R_{1/2}$), $1.0 < R_{1/2} < 5.0 \ \rm arcsec$; and the concentration index ($C$, \citealt{Conselice2000}), $2.1 < C < 5$; 
and extinction corrected colour constraints: $0.5<(J - K_{s})< 2.0 \ \rm mag$; 
$0.0 < (J - H) < 1.0 \ \rm mag$;  
$0.0 < (H - K_{s}) < 2.0 \ \rm mag$; and 
$(J - H) + 0.9 (H - K_{s}) > 0.44 \ \rm mag$ to separate point and extended sources.   
Finally, to eliminate false detections, they performed a visual inspection 
to confirm all the galaxies \citep[see][for more details]{Baravalle2021}.

It is important to note that only 45 galaxies were previously studied by other authors in these regions: 8 of them have radial velocity estimates and one was reported as radio galaxy CenB \citep{West1989}, also known as 2MASX J13464910-6024299 or WISEA J134649.05-602429.3.  Nine sources are in Gaia-DR3 \citep{Brown2021} and were eliminated in subsequent studies involving the catalogue \citep{Soto2022}. Therefore, we considered in this analysis 5554 galaxies from the VVV NIRGC. We used the interstellar extinction corrected total magnitudes in the $J$, $H$ and $K_{s}$ passbands and aperture colours ($J$-$K_{s}$), ($J$-$H$) and ($H$-$K_{s}$) obtained from magnitudes within a fixed aperture of 2 arcsec.

In addition, the WISE survey mapped the whole sky at 3.4, 4.6, 12, and 22 $\mu$m 
passbands with an angular resolution of 6.1, 6.4, 6.5 and 12.0 arcsec in the four passbands, respectively \citep{Wright2010}. 
Here, we use the  AllWISE Source Catalogue that contains the photometry in the four passbands of the detected objects 
and we use the mid-IR magnitudes to estimate the colours (3.4-4.6), (4.6-12) and (12-22). The VVV NIRGC includes 509 galaxies in common with the WISE survey with angular separations of less than 2 arcsec \citep{Baravalle2021}.   
 
On the other hand, \cite{Molnar2021} applied an automated algorithm to obtain the variability classification of the VVV data generating the VIVACE catalogue.  
This catalogue contains the main statistical properties of the  
light-curves used for the classification as well as the mean $Z$, $Y$, $J$, $H$, $K_{s}$ magnitudes and periods together with a cross-match with Gaia-EDR3 \citep{Brown2021}.

\subsection{The final sample}
\label{sample}

Our sample contains galaxies from the VVV NIRGC with available mid-IR data from WISE. Using angular separations lower than 1 arcsec, we have five of these galaxies in the VIVACE catalogue as variable sources. 
These objects have remained unexplored and they are studied and classified here for the first time.  
Table~\ref{table1} shows the identifications and coordinates of the five studied objects.  

\begin{figure*}
\begin{centering}
    \includegraphics [width=\textwidth]{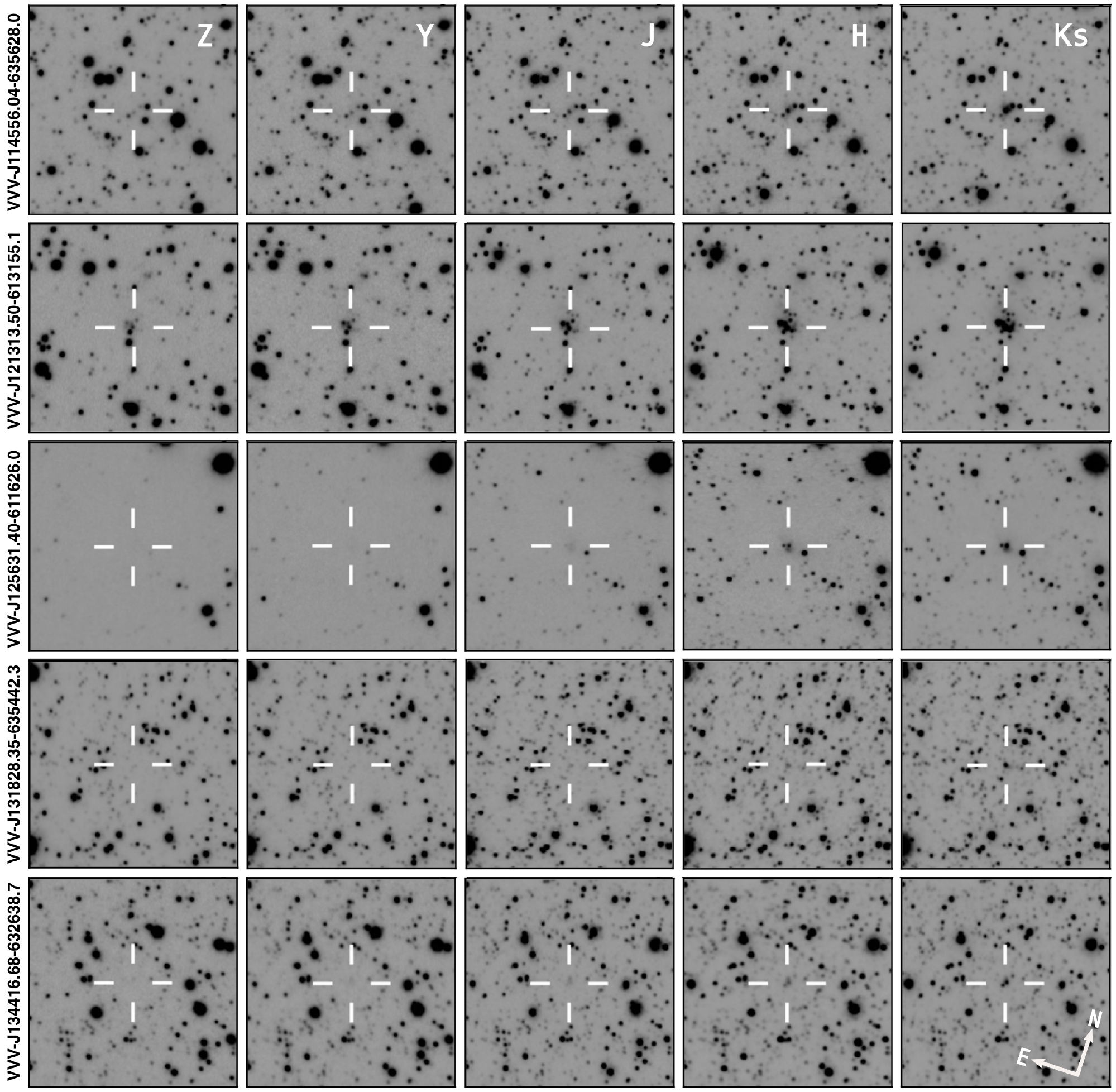}
  \caption{1$^{\prime}$ $\times$ 1$^{\prime}$ VVV images of the five studied
    sources in the $Z$, $Y$, $J$, $H$ and $K_{s}$ passbands.  The orientation of all images is shown in the bottom-right panel.}
	\label{f1}
\end{centering}
\end{figure*}

\begin{figure*}
\begin{centering}
  \includegraphics [scale=0.8]{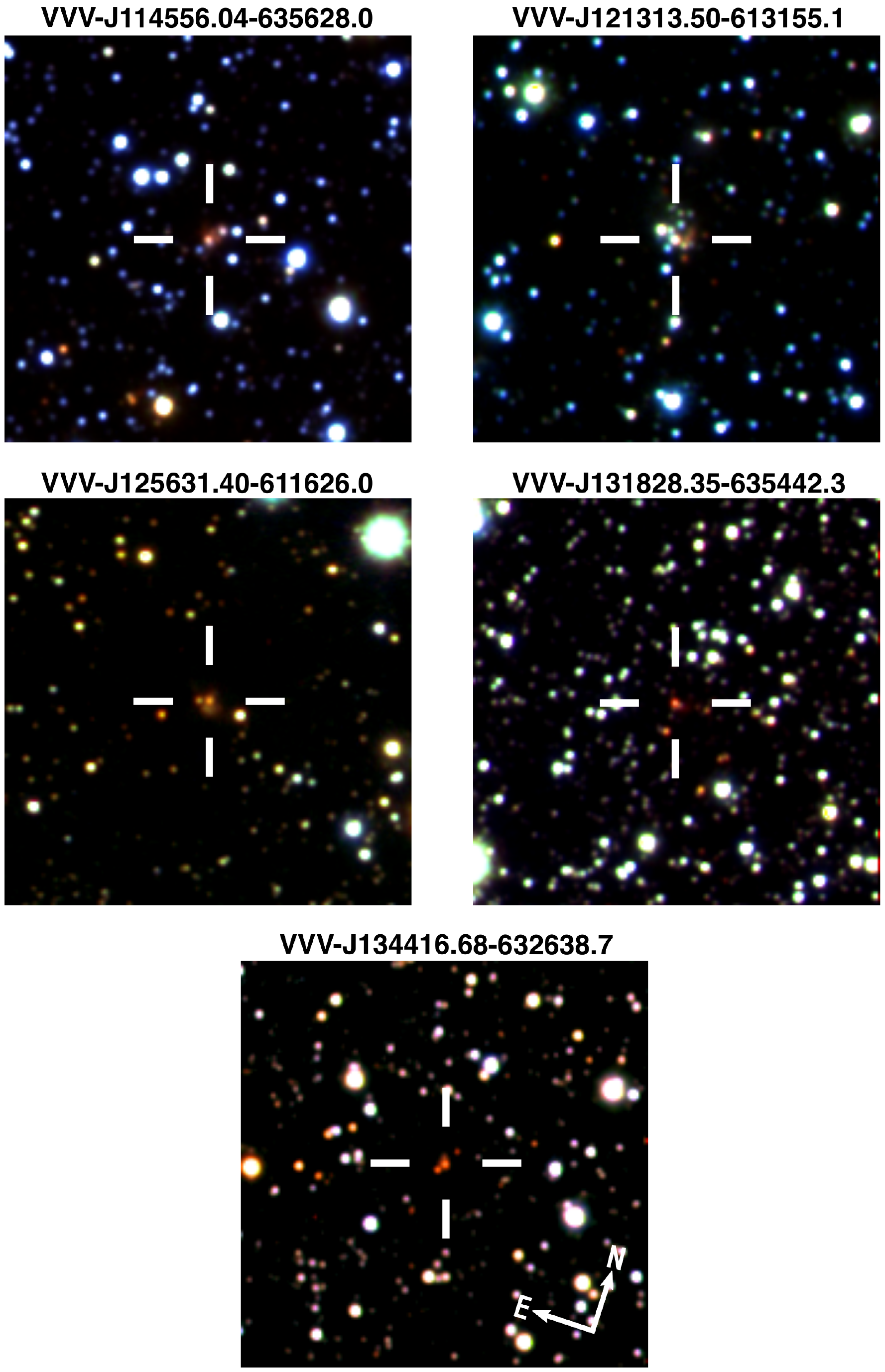}
  \caption{1$^{\prime}$ $\times$ 1$^{\prime}$ VVV colour composed images of the five studied
    sources. The orientation of all images is shown in the bottom panel.}
	\label{f2}
\end{centering}
\end{figure*}

Among our objects of study, only the galaxy VVV-J114556.04-635628.0 was previously identified as the source IRAS 11435-6339 \citep{Cutri2003}.
On the other hand, the only known radio galaxy in the VVV NIRGC is CenB, which is not studied as variable source in the VIVACE catalogue and therefore is not included in this work.
Figure~\ref{f1} shows the 1$^\prime \times$ 1$^\prime$ images in the $Z$, $Y$, $J$, $H$ and $K_{s}$ passbands of the VVV survey for the five studied objects. These objects present typical behaviour of extragalactic sources in the different near-IR passbands, stronger fluxes in the $K_{s}$ passband decreasing towards $H$, $J$, $Y$ and $Z$ passbands as described in \cite{Baravalle2018}.
Figure~\ref{f2} shows the 1$^\prime \times$ 1$^\prime$ colour composed images for these five sources. They are morphologically distinct and redder than the surrounding point-source-like stars. As we will see below, the analysis of the colour-colour diagrams will give us some clues of the nature of these objects.

In this analysis, it is important to use the ($Y$-$J$) colour to discriminate among QSOs \citep{Cioni2013}.
We also obtained the photometry for the $Y$ images using our photometric procedure described in \cite{Baravalle2018} with \se + PSFEx.
As seen in Figure~\ref{f1}, the $Z$ and $Y$ brightness of the objects are in general fainter than the other passbands. The VVV-J114556.04-635628.0 and VVV-J121313.50-613155.1 are also contaminated by nearby sources. There are only two cases where we could obtain the $Y$ magnitudes with smaller uncertainties. In Table~\ref{table2} we present the near-IR magnitudes of the studied sources taken from the VVV NIRGC. As mentioned above, the $Y$ magnitude was obtained in this work. 
The range of $K_{s}$ total magnitudes is from 12.25 to 13.88 mag. The object VVV-J125631.40-611626.0 is the brightest of the sample and has the highest interstellar extinction as clearly seen in Figure~\ref{f1}.

Table~\ref{table3} shows the mid-IR magnitudes taken from WISE.
In general there are four sources with colours (3.4 - 4.6) $<$ 1.3 mag and (4.6-12) $>$ 3.5 mag with the exception of  VVV-J125631.40-611626.0, that presents very different colours.   

We also checked these five sources in the Gaia-DR3 \citep{GAIADR3}
using an angular separation of 2 arcsec between sources with the restriction of the Renormalised Unit Weight Error (RUWE) smaller than 1.4. The source VVV-J121313.50-613155.1 is the only one with Gaia counterparts. Two sources were found in the field, the closest one, 6057779246181521792, with an angular separation of 0.37 arcsec, has parallax of (0.87 $\pm$ 1.41) mas, which is not statistically significant as a positive parallax. The visual inspection for this source was crucial because it is placed in a very crowded region (Fig.~\ref{f2}). 

\begin{table*}
\center
\begin{tabular}{|ccccccc|}
\hline 
  VVV NIRGC &  VIVACE         & R.A.     & Decl. & Galactic longitude & Galactic latitude & Tile ID\\ 
  ID         &    ID          &  (J2000) & (J2000) & [degree]     & [degree] & \\
\hline 
 VVV-J114556.04-635628.0 & 1363656 & 11:45:56.04 & -63:56:28.0 & 295.813 & -1.978 & d001\\
 VVV-J121313.50-613155.1 & 1081363 & 12:13:13.50 & -61:31:55.1 & 298.394 & 1.002  & d079\\
 VVV-J125631.40-611626.0 & 1068365 & 12:56:31.40 & -61:16:26.0 & 303.543 & 1.592  & d120\\ 
 VVV-J131828.35-635442.3 & 1338147 & 13:18:28.35 & -63:54:42.3 & 305.899 & -1.196 & d008\\
 VVV-J134416.68-632638.7 &   * & 13:44:16.68 & -63:26:38.7 & 308.797 & -1.176 & d010\\
 \hline 
 \multicolumn{3}{l}{* D.Minniti. (priv. comm)}\\
\end{tabular}
\caption{Studied sources in this work. 
The VVV NIRGC and VIVACE identifications of the five studied objects are listing in columns (1) and (2), respectively; in columns (3) to (6), the equatorial and Galactic coordinates; and in column (7), the VVV tile identifier where the objects are located.}
\label{table1}
\end{table*}

\begin{table*}
\center
\begin{tabular}{cccccccccc}
\hline 
  VVV NIRGC   & A$K_{s}$    & Y     &    J   &   H    & K$_{s}$ & (J-K$_{s}$) & (J-H) & (H-K$_{s}$) & (Y-J) \\
   ID         & [mag]      & [mag] & [mag]  &  [mag] & [mag]   & [mag]       & [mag] & [mag]       & [mag]\\
\hline 
VVV-J114556.04-635628.0 & 0.87 &  --   & 15.34$\pm$0.03 & 14.02$\pm$0.01 & 13.36$\pm$0.02  & 1.43$\pm$0.05 & 0.82$\pm$0.03 & 0.63$\pm$0.03 & --  \\
VVV-J121313.50-613155.1 & 0.60 &  --   & 14.42$\pm$0.01 & 13.32$\pm$0.01 & 12.57$\pm$0.01  & 1.69$\pm$0.02 &  0.96$\pm$0.02 & 0.73$\pm$0.01 & --  \\
VVV-J125631.40-611626.0 & 1.93  & --   & 13.08$\pm$0.05 & 12.54$\pm$0.02 & 12.25$\pm$0.01  & 1.36$\pm$0.06 & 0.82$\pm$0.06 & 0.54$\pm$0.02 & -- \\
VVV-J131828.35-635442.3 & 0.82 & 16.22$\pm$0.10 & 15.76$\pm$0.07 & 14.96$\pm$0.04 & 13.88$\pm$0.02  & 1.89$\pm$0.09 &  0.88$\pm$0.09 & 1.01$\pm$0.04 & 0.29$\pm$0.17\\
VVV-J134416.68-632638.7 & 1.51 & 14.46$\pm$0.09 & 13.98$\pm$0.06 & 13.46$\pm$0.03 & 13.12$\pm$0.02  & 1.07$\pm$0.09 & 0.57$\pm$0.08 & 0.50$\pm$0.04 & 0.22$\pm$0.15 \\
\hline 
\end{tabular}
\caption{Near-IR photometry of the studied sources. Column (1) shows the VVV NIRGC identifications; column (2), the AK$_{s}$ interstellar extinction; columns (3) to (6), the $Y$, $J$, $H$ and $K_{s}$ total magnitudes; and in columns (7) to (10), the ($J$-$K_{s}$), ($J$-$H$), ($H$-$K_{s}$) and ($Y$-$J$) colours obtained from magnitudes within a fixed aperture of 2 arcsec diameter.}
\label{table2}
\end{table*}

\begin{table*}
\center
\begin{tabular}{ccccccc}
\hline 
 VVV NIRGC   & [3.4]  & [4.6] & [12]  & [22]  & (3.4 - 4.6) & (4.6 - 12)\\ 
     ID     &  [mag] & [mag] & [mag] & [mag] & [mag] & [mag]\\
\hline 
VVV-J114556.04-635628.0  & 11.55$\pm$0.04 & 10.77$\pm$0.02 & 6.01$\pm$0.01 & 3.60$\pm$0.02 & 0.78$\pm$0.06 & 4.76$\pm$0.03\\
VVV-J121313.50-613155.1  & 11.10$\pm$0.03 & 9.86$\pm$0.02  & 6.33$\pm$0.02 & 2.54$\pm$0.02 & 1.24$\pm$0.05 & 3.53$\pm$0.04\\
VVV-J125631.40-611626.0 & 12.18$\pm$0.03 & 10.62$\pm$0.02 & 8.54$\pm$0.03 & 5.04$\pm$0.03 & 1.56$\pm$0.05 & 2.08$\pm$0.05\\
VVV-J131828.35-635442.3 & 12.59$\pm$0.05 & 11.66$\pm$0.04 & 6.91$\pm$0.02 & 4.95$\pm$0.04 & 0.93$\pm$0.09 & 4.75$\pm$0.06\\
VVV-J134416.68-632638.7   & 12.89$\pm$0.06 & 11.91$\pm$0.03 & 7.70$\pm$0.03 & 4.37$\pm$0.04 & 0.98$\pm$0.09 & 4.21$\pm$0.06 \\
\hline 
\end{tabular}
\caption{Mid-IR photometry of the studied sources. The VVV NIRGC identifications are listed in column (1); the WISE magnitudes in columns (2) to (5); and the WISE (3.4 - 4.6) and (4.6-12) colours in columns (6) and (7), respectively.}
\label{table3}
\end{table*}

\subsection{Colour-colour diagrams}

\begin{figure*}
   \includegraphics [scale=0.8]{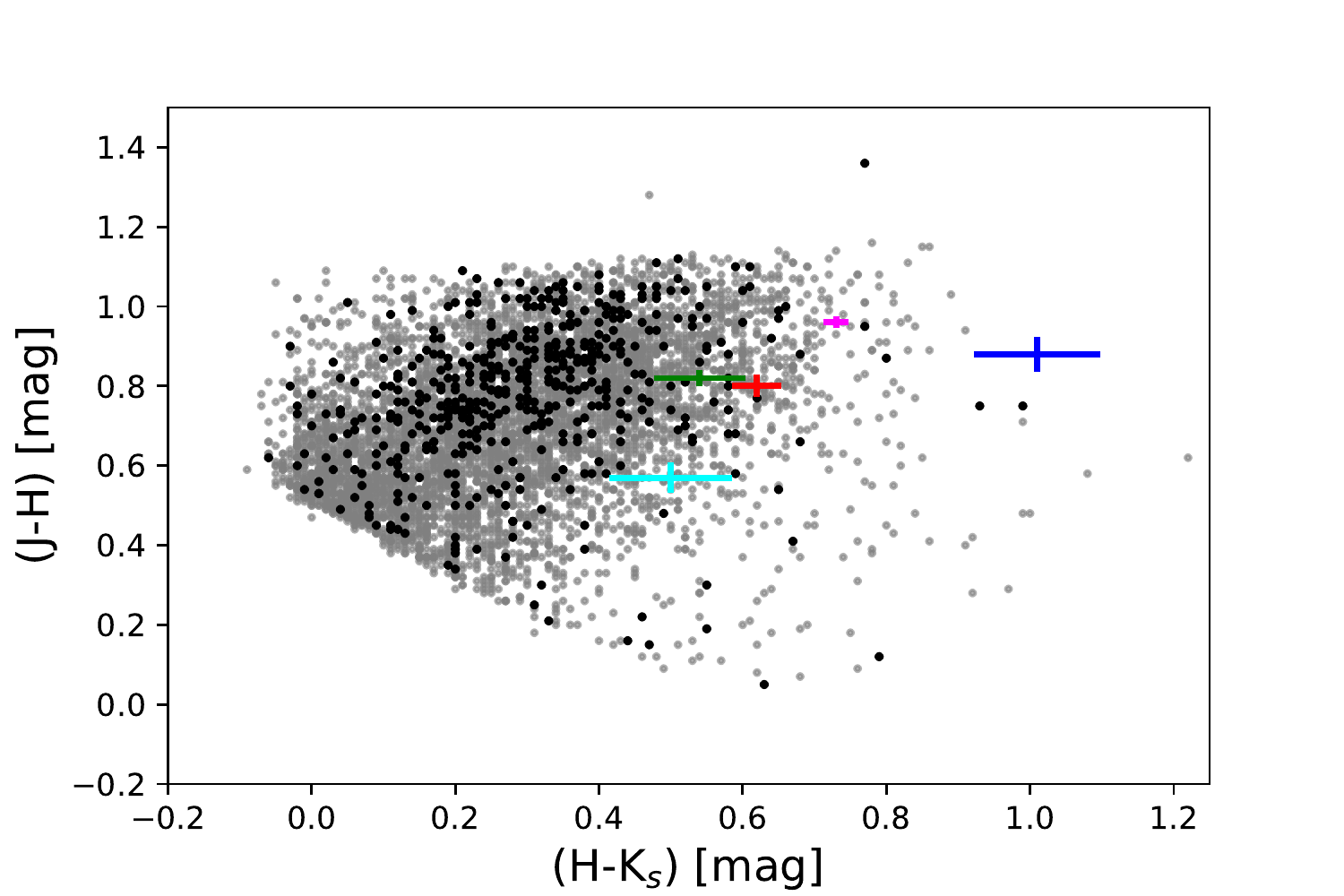}
\caption{Near-IR colour–colour diagram from the VVV survey. The VVV NIRGC galaxies are represented by grey points and those galaxies with WISE counterpart, with black circles. The five studied sources in this work are identified with different colours: VVV-J114556.04-635628.0 in red; VVV-J121313.50-613155.1 in pink; VVV-J125631.40-611626.0 in green; VVV-J131828.35-635442.3 in blue and VVV-J134416.68-632638.7 in cyan. Error bars are considered at 1 $\sigma$.}
\label{fig:diagramasVVV}
\end{figure*}

When searching in large near-IR databases for AGNs at lower latitudes, we must also consider young stellar objects (YSOs) as contaminants. Related to this, typical YSOs have very red colours \citep[e.g.,][]{Lucas2017, Medina2021}.
The colour-colour diagrams are a powerful tool for separating point (stellar) from extragalactic sources
(eg. \citealt{2000AJ....119.2498J, Baravalle2018}). 
In particular, using near-IR colours it is also possible to identify AGNs from these extragalactic sources (eg. \citealt{Jarrett2011,Stern2012,Cioni2013,Jarret2017}). 
Figure~\ref{fig:diagramasVVV} shows the ($J$ - $H$) vs ($H$ - $K_{s}$) colour-colour diagram using near-IR data from the VVV survey.
The VVV NIRGC galaxies are represented by grey points and the galaxies with WISE counterparts are also highlighted by black circles. 
The five studied sources in this work are identified with different colours and the error bars at 1 $\sigma$ are also included.
Our objects present values ($H$ - $K_{s}$) > 0.5 mag and ($J$ - $H$) > 0.55 mag, being in general redder than most of the remaining sources (see also Table~\ref{table2}).

\cite{Cioni2013} identified QSOs behind the Magellanic system using the near-IR ($Y$ - $J$) vs ($J$ - $K_{s}$) colour-colour diagram using the VISTA survey of the Magellanic Clouds.  
In their Figure 2, they defined two regions were the QSOs are located. 
The relation ($J$ - $K_{s}$) = -1.25 $\times$ ($Y$ - $J$) + 1.90 marks the division between star-like and galaxy-like QSOs. Unfortunately, we have only two galaxies
of our sample with accurate $Y$ magnitudes. The other three present strong contamination or too faint $Y$ surface brightness (Figure~\ref{f1}). The ($Y$ - $J$) colours are reported in Table~\ref{table2}. Following \cite{Cioni2013}, the galaxy VVV-J131828.35-635442.3 is located in the galaxy-type QSO region and VVV-J134416.68-632638.7 in the star-like QSO region. 

Moreover, in the mid-IR passband \cite{Stern2012} used the colour criterion (3.4 - 4.6) $>$ 0.8 mag to detect AGNs. Following this, \cite{Assef2018} presented other two colour criteria: (3.4 - 4.6) $>$ 0.5 mag and (3.4 - 4.6) $>$ 0.77 mag in their AGN sample. 
\cite{Koenig2012} used the mid-IR (3.4 - 4.6) vs (4.6 - 12) 
colour-colour diagram with WISE data to separate YSOs and AGNs. In this context, the source VVV-J125631.40-611626.0 has different mid-IR colours compared to the other four sources.

\begin{figure*}
\begin{center}
\includegraphics[scale=0.8]{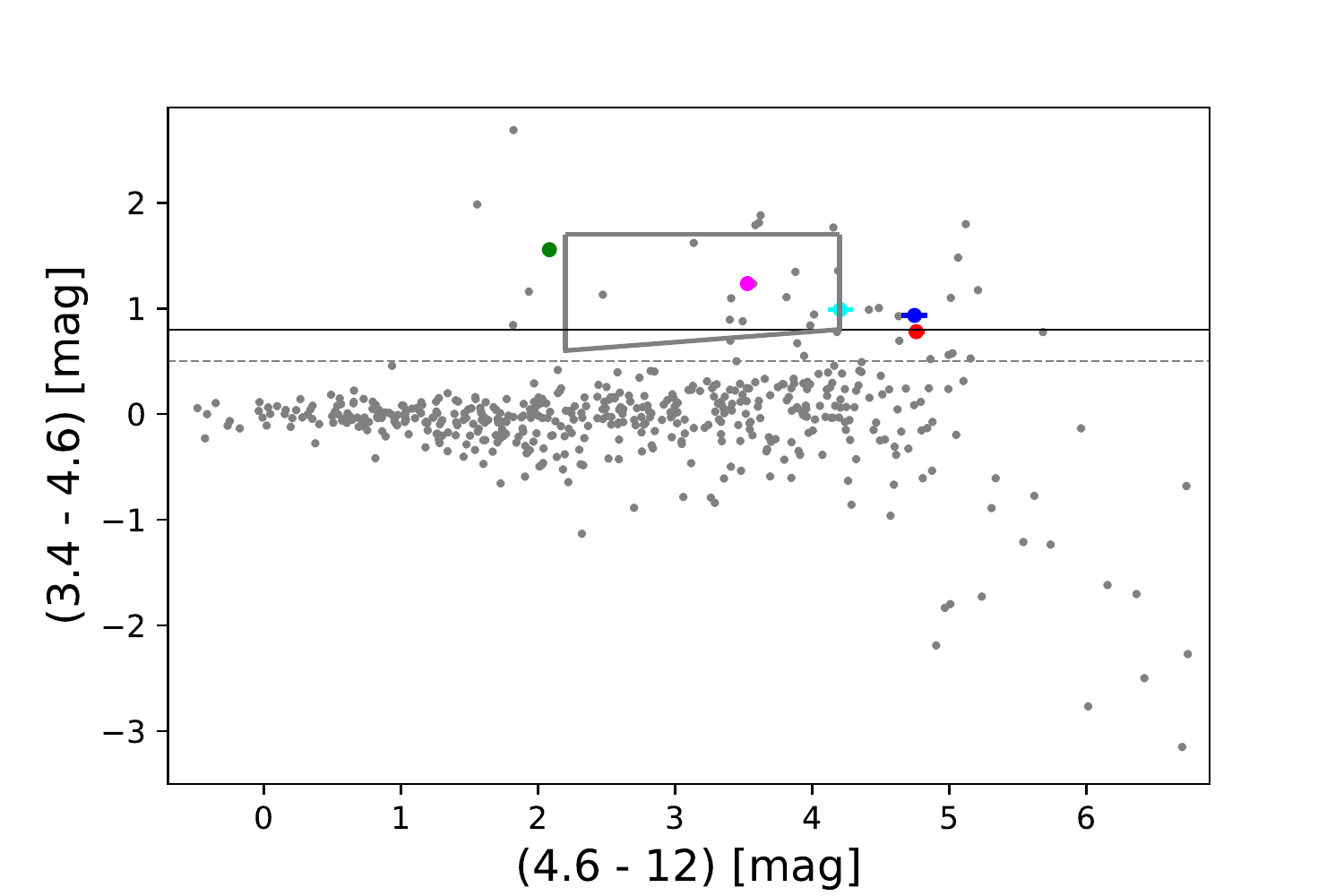}
\end{center}
\caption{Mid-IR colour–colour diagram for galaxies in common with the VVV NIRGC and WISE. The five studied sources in this work are represented with the same colours as previous figure: VVV-J114556.04-635628.0 in red; VVV-J121313.50-613155.1 in pink; VVV-J125631.40-611626.0 in green; VVVJ131828.35-635442.3 in blue and VVV-J134416.68-632638.7 in cyan. Error bars are considered at 1 $\sigma$. The solid black and dashed gray horizontal lines represent the limits for AGNs from \citet{Stern2012} and \citet{Assef2018}, respectively. The gray box denotes the defined region of QSOs/AGNs from \citet{Jarrett2011}. }
\label{diagramaWISE}
\end{figure*}

Figure~\ref{diagramaWISE} shows the WISE mid-IR colour-colour diagrams of the 
WISE sources in common with the VVV NIRGC galaxies.  
The five sources are shown on the figure using the same colour notation as Figure~\ref{fig:diagramasVVV} and the error bars are at 1 $\sigma$. 
The black and blue horizontal dashed lines are the AGN threshold from \cite{Stern2012} and \cite{Assef2018}, respectively. The grey box denotes the region defined by \cite{Jarrett2011} containing AGNs where in general the QSOs and Seyfert galaxies are located. 
According to our results, the five sources satisfy both the \cite{Assef2018} and also, within the uncertainties, the \cite{Stern2012} criteria to be QSO candidates. 
On the other hand, galaxies VVV-J121313.50-613155.1 and VVV-J134416.68-632638.7 are located inside the box defined by \cite{Jarrett2011}. 
In addition, the source VVV-J125631.40-611626.0 is located outside the regions populated by galaxies
showed in \cite{Wright2010} (their Figure 12) but in the YSO region defined by \citealt{Koenig2012} (their Figure 7).

\begin{figure*}
\includegraphics[width=0.48\textwidth]{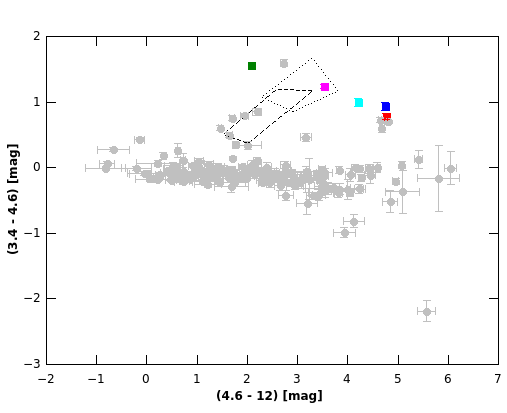}
\includegraphics[width=0.50\textwidth]{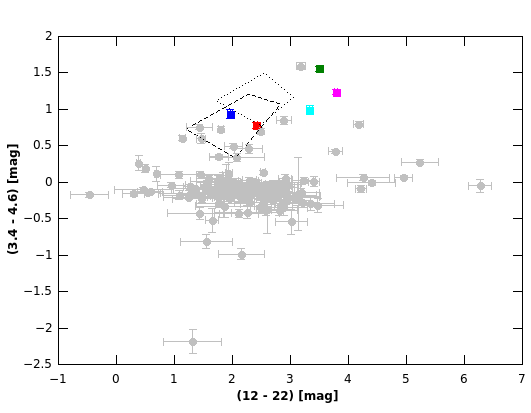}
\caption{Mid-IR colour-colour diagrams using WISE. The two boxes corresponding to the two blazar classes of BZBs (BL Lac) and BZQs (FSRQ) are shown in dash- and dot- black lines. All the WISE objects are represented by grey points and the five studied sources in this work with the same colours as previous figures: VVV-J114556.04-635628.0 in red; VVV-J121313.50-613155.1 in pink; VVV-J125631.40-611626.0 in green; VVVJ131828.35-635442.3 in blue and VVV-J134416.68-632638.7 in cyan.}
\label{CCD1}
\end{figure*}

Using the mid-IR (3.4 - 4.6) vs (12 - 22) colour-colour diagram, \cite{DAbrusco2012} defined a two dimensional region, the WISE Gamma-ray Strip (WGS) where the blazars are located. \cite{Pichel2020} using both, VVV and WISE data, applied the WGS method showing that blazars might populate specific regions in the near- and mid-IR colour-colour diagrams, clearly separated from other extragalactic sources. Following \cite{Pichel2020}, we performed the WGS test in order to examine the location of the galaxies in the bi-dimensional region defined in the mid-IR colour-colour diagrams. Using the positions of the five sources, we selected all the WISE objects within a circle of 2 arcmin of the source. Figure~\ref{CCD1} shows the mid-IR (3.4 - 4.6) vs (4.6 - 12) and (3.4 - 4.6) vs (12 - 22) colour-colour diagrams of all the WISE sources (represented by grey points)
in the regions of these five VVV NIRGC sources (represented by the same colour notation as Figure~\ref{fig:diagramasVVV}). 
The two boxes corresponding to the two blazar classes of BZBs (BL Lac) and BZQs (FSRQ) are shown in dash- and dot- black lines, respectively.  
The objects  VVV-J114556.04-635628.0, VVV-J121313.50-613155.1 and VVV-J131828.35-635442.3 satisfy only one of the criteria to be considered blazars found by \cite{DAbrusco2019} and references therein.

\section{Variability analysis}
\label{sec:variability}

\begin{figure*}
\begin{center}
\includegraphics[width=0.46\textwidth]{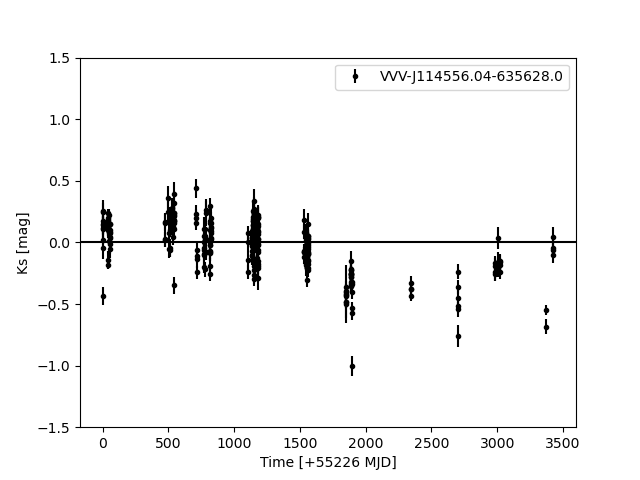}
\includegraphics[width=0.46\textwidth]{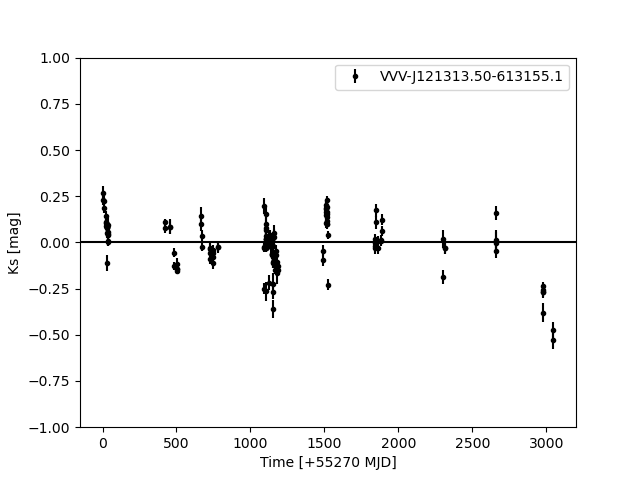}\\
\includegraphics[width=0.46\textwidth]{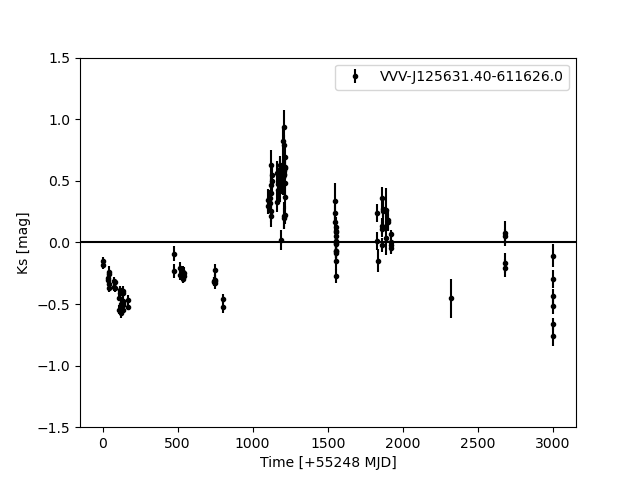}
\includegraphics[width=0.46\textwidth]{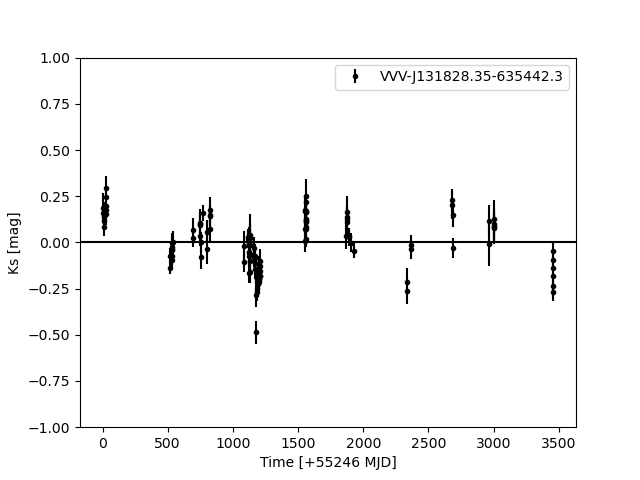}\\
\includegraphics[width=0.46\textwidth]{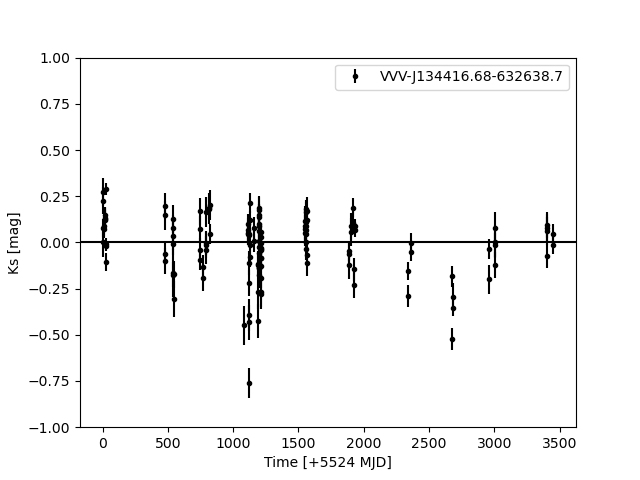}
\caption{$K_{s}$ differential light curves of the five studied sources.}
\label{fig:lightcurves}
\end{center}
\end{figure*}

We obtained the profile fitted $K_{s}$ band light curves from which the VIVACE sources were originally classified \citep[see][for a description of these data]{Molnar2021}.
The differential $K_{s}$ light curves of the five studied sources are shown in Figure~\ref{fig:lightcurves}. 
These are median subtracted curves sampled over more than 3000 days. 
The shape of the light curves is, in general, irregular without any clear periodicity. 
In some cases, peak-shaped changes in brightness are observed, which are significant with respect to the uncertainties.  

In order to analyse the variability of the objects, we applied two different statistical methods.
First, we studied the fractional variability amplitude $\sigma_{rms}$, defined as:

\begin{equation}
\label{eq:sigma}
 \sigma_{rms}^2=  \frac{1}{N \mu^2}  \sum_{i=1}^{N}[ (F_{i}-\mu)^2 - \epsilon_{i}^2 ],
\end{equation}

where N is the number of flux values $F_{i}$, with measurement uncertainties $\epsilon_{i}$, and $\mu$ is the average flux. This parameter is widely used to evaluate and analyse variability in light curves \citep[e.g.,][]{Nandra1997, Edelson2002, Sandrinelli2014, Pichel2020}, representing an excess of variability that cannot be only generated by flux errors.

Secondly, we analysed the slope of the light curves taking into account that \cite{Cioni2013} showed that 75\% of the QSOs have a slope variation in the $K_{s}$ passband $>10^{-4}$ mag/day. These authors derived the slope of the overall $K_{s}$ variation in light curves sampled over 300$-$600 days, 40$-$80 days or shorter range. In order to carry out this analysis, we performed a linear fit of the $K_{s}$ light curves considering a range of days defined by the highest and lowest variations. In all cases, the time range considered is longer than 1200 days. 

Table~\ref{table4} shows the $K_{s}$ variability of the five studied sources.
As mentioned above, the shape of the light curves is in certain cases irregular and they do not show any clear periodicity. The VIVACE catalogue contains some mis-classifications of aperiodic variable sources as they were not included in the training set and the classification system. Indeed, \cite{Molnar2021} noted that aperiodic YSOs and/or AGNs, for example, might contaminate the long period variable category. 
However, for completeness, we added in the table the periodicity from VIVACE, which is a statistical parameter \citep[see][for more details]{Molnar2021}. 
We visually inspected the light curves phase-folded with the VIVACE period and did not find them to be particularly convincing fits to these data.

\begin{table*}
\center
\begin{tabular}{|ccccc|}
\hline 
 VVV NIRGC             & Period &    $\sigma_{rms}$    & Range of days  &  Slope variation   \\ 
     ID                &  [day] &   ($\times$ 100\%)   &    [day]       &  [mag/day]         \\
\hline 
VVV-J114556.04-635628.0 & 140  & 20.7                               & 400 - 2000  & 0.00031                 \\
VVV-J121313.50-613155.1 & 140  & 13.3                               & 1800 - 3100 & 0.00028                 \\
VVV-J125631.40-611626.0 & 1364 & 33.8                               & 1100 - 3100 & 0.00047                 \\ 
VVV-J131828.35-635442.3 & 553  & 12.6                               & 0 - 1200    & 0.00026                 \\
VVV-J134416.68-632638.7 & 209  & 16.1                               & 1500 - 2700 & 0.00032                 \\
\hline 
\end{tabular}
\caption{$K_{s}$ variability of the studied sources. The VVV NIRGC identification of the studied objects are listed in column (1); the period listed in the VIVACE catalogue in column (2); the fractional variability amplitude ($\times$ 100\%) calculated with Eq.~\ref{eq:sigma} in column (3); the range of days considered to derive the slope in column (4) and the absolute value of the slope of the variation in column (5).}
\label{table4}
\end{table*}

\begin{table*}
\center
\begin{tabular}{|cccccccc|}
\hline 
 VVV NIRGC & \citeauthor{Assef2018}  & \citeauthor{Stern2012} & \citeauthor{Jarrett2011}  & \citeauthor{Wright2010} & \citeauthor{Koenig2012} & \multicolumn{2}{c}{\cite{Cioni2013}} \\ 
     ID    & \citeyearpar{Assef2018} & \citeyearpar{Stern2012}& \citeyearpar{Jarrett2011} & \citeyearpar{Wright2010}& \citeyearpar{Koenig2012} & ($Y$ - $J$) & slope \\
\hline 
VVV-J114556.04-635628.0 &  \checkmark & \checkmark &            & \checkmark & \checkmark &     *    & \checkmark \\
VVV-J121313.50-613155.1 &  \checkmark & \checkmark & \checkmark & \checkmark & \checkmark &     *    & \checkmark\\
VVV-J125631.40-611626.0 &  \checkmark & \checkmark &            &            &            &     *    & \checkmark\\ 
VVV-J131828.35-635442.3 &  \checkmark & \checkmark &            & \checkmark & \checkmark & \checkmark & \checkmark\\
VVV-J134416.68-632638.7 &  \checkmark & \checkmark & \checkmark & \checkmark & \checkmark & \checkmark & \checkmark\\
\hline
 \multicolumn{3}{l}{* No colour available (see Sect.~\ref{sample}).}\\
\end{tabular}
\caption{Summary of the different criteria satisfied by the five studied objects to be QSO/AGN candidates.}
\label{table5}
\end{table*}

\section{Discussion and final remarks}
\label{sec:final}

In this work we studied the five sources from the VVV NIRGC that also have WISE counterparts and present variability in the VIVACE catalogue.  
In the near-IR colour-colour diagrams, our objects present values ($H$ - $K_{s}$) $>$ 0.5 mag
and ($J$ - $H$) $>$ 0.55 mag, being in general redder than most of the remaining sources 
(Fig.~\ref{fig:diagramasVVV}).  
The two galaxies with $Y$ photometric data (VVV-J131828.35-635442.3 and VVV-J134416.68-632638.7) are both located in the QSO region following \cite{Cioni2013}.

Analysing the mid-IR colour-colour diagram  (Fig.~\ref{diagramaWISE}), we found that the five sources satisfy both the \cite{Assef2018} and \cite{Stern2012} criteria, suggesting that our sources could be AGNs. In addition, the objects VVV-J121313.50-613155.1 and VVV-J134416.68-632638.7 
are located in the AGN region defined empirically by \cite{Jarrett2011}. 
The object VVV-J125631.40-611626.0 is in the YSO region of the mid-IR colour-colour diagram of \cite{Koenig2012} and also it is not located in the region populated by galaxies in \cite{Wright2010}.
Using the WGS method, from Figure~\ref{CCD1} the WISE sources coincident with VVV-J114556.04-635628.0, VVV-J121313.50-613155.1 and VVV-J131828.35-635442.3 satisfy only one of the criteria to be a blazar candidate. The sources VVV-J125631.40-611626.0 and VVV-J134416.68-632638.7 do not satisfy any of the criteria. Based on this, none of the objects present sufficient evidence to be blazar candidates.

On the other hand, we analysed the variability using two different statistical methods: the fractional variability
amplitude $\sigma_{rms}$ (Eq.~\ref{eq:sigma}) and the slope of the light curve (see Sect.~\ref{sec:variability}). 
The values of $\sigma_{rms}$ are in the range 12.6 $-$ 33.8, which are in accordance to those found in type-1 AGNs \citep[e.g.,][]{Nandra1997, Edelson2002}. 
These values are, however, lower than those found in blazars \citep[e.g.,][]{Sandrinelli2014, Pichel2020}. Considering that in general type-1 AGNs present smaller variability amplitudes than blazars \citep[e.g.,][]{Mao2021}, these results suggest that these objects are type-1 AGN candidates; i.e. quasars and/or Seyfert 1 galaxies. Additionally, the slopes of the light curves, which are in the range (2.6$-$4.7) $\times 10^{-4}$ mag/day, are well within the limit value ($>$ 1 $\times 10^{-4}$ mag/day) reported by \cite{Cioni2013} for quasars.

The object VVV-J125631.40-611626.0 presents the highest $\sigma_{rms}$ value and following \cite{Lucas2017} the YSOs show the highest variations in the $K_{s}$ light curves. In this sense, typical YSOs exhibit variability that is stochastic, bursting or semi-periodic in short or long time scales. Indeed, these objects have been previously characterised in detail in the VVV survey (e.g., \citealt{Lucas2017,Contreras2017,Medina2021,Guo2020,Guo2022}).

In summary, Table~\ref{table5} shows the various observational evidences for each studied source in favour or against the criteria used by different authors to be considered AGN candidates.
All these results suggest that four sources could be type-1 AGNs and the source VVV-J125631.40-611626.0 might be a YSO.

This kind of work allow us to find AGNs at lower Galactic latitudes where there is a lack of these studies. 
Future data coming from Vera C. Rubin Observatory Legacy Survey of Space and Time \citep{LSST2009} and the eROSITA X-ray telescope \citep{Brunner2022}
would complement the VVV and VVVX databases. This synergy among data coming from different wavelengths  allow us to deeply investigate these interesting galaxies. Further work involving spectroscopy is also crucial to shed light on their nature and active properties.

\section*{Acknowledgements}

We would like to thank the anonymous referee for the useful comments and suggestions which has helped to improve this paper.
This work was partially supported by Consejo de Investigaciones Cient\'ificas y T\'ecnicas (CONICET) and Secretar\'ia de Ciencia y T\'ecnica de la Universidad Nacional de C\'ordoba (SecyT). 
The authors gratefully acknowledge data from the ESO Public Survey program IDs 179.B-2002 and 198.B-2004 taken with the VISTA telescope, and products from the Cambridge Astronomical Survey Unit (CASU).
D.M. gratefully acknowledges support by the ANID BASAL projects ACE210002 and FB210003 and by Fondecyt Project No. 1220724. 

\section*{Data Availability}

The VVV NIRGC used in this article is available at
\url{https://catalogs.oac.uncor.edu/vvv\_nirgc/}. 
The VIVACE catalogue is available at \url{https://people.ast.cam.ac.uk/~jls/data/vproject/vivace_catalogue.fits}.





\bibliographystyle{mnras}
\bibliography{source.bib} 








\bsp	
\label{lastpage}
\end{document}